# Horizontal gene transfer drives extreme physiological change in Haloarchaea.

*Christopher J. Creevey[1] and James O. McInerney[2]*

Current author affiliations:

[1] *Animal & Bioscience Research Department, Teagasc, Grange, Co. Meath, Ireland..*

[2] *Department of Biology, National University of Ireland Maynooth, Co. Kildare, Ireland.*

**Corresponding author:**
Dr. Chris Creevey
Animal & Bioscience Research Department,
Teagasc,
Grange,
Co. Meath,
Ireland.

Keywords: Horizontal gene transfer, Physiology, Genomics, Archaea.

Running title: The bacterial ancestry of the Archaea.

**The haloarchaea are aerobic, heterotrophic, photophosphorylating prokaryotes, whose supposed closest relatives and ancestors, the methanogens, are $CO_2$-reducing, anaerobic chemolithotrophs. Using two available haloarchaeal genomes we firstly confirmed the methanogenic ancestry of the group and then investigated those individual genes in the haloarchaea that differ in their phylogenetic signal to this relationship. We found that almost half the genes, about which we can make strong statements, have bacterial ancestry and are likely a result of multiple horizontal transfer events. Futhermore their functions specifically relate to the phenotypic changes required for a chemolithotroph to become a heterotroph. If this phylogenetic relationship is correct, it implies the development of the haloarchaeal phenotype was among the most extreme changes in cellular physiology fuelled by horizontal gene transfer.**

# Introduction:

Within the Euryarchaea, the haloarchaea are considered to be a monophyletic group [1] and due to their mesophilic, heterotrophic, aerobic and photophosphorylative nature, their phylogenetic affiliation with the other Archaea was not realised initially [2]. Some strains of haloarchaea have bacteriorhodopsin bilayer membrane components that serve as light-driven proton pumps. Light energy is used to pump protons out of the cell and thereby generate an electrochemical potential. This, in turn drives ATP synthesis. This is similar in nature to the mechanism used by bacterial proteorhodopsin. The haloarchaea are also aerobic and heterotrophic and the ease with which they will grow under laboratory conditions is part of the reason they have been suggested to be ideal candidates for the study of archaeal systems biology [1, 3].

Phylogenies constructed from both Small subunit (SSU) ribosomal RNA phylogenies [4] and universally distributed protein coding genes [5] place the haloarchaea as a sister group to the Methanomicrobiales, deep within the methanogenic Archaea suggesting that they are derived from a methanogenic ancestor [1]. Energy production for the haloarchaea is significantly different to the Methanomicrobiales, whereas gene structure, DNA replication, transcription and translation as well as lipid linkages are very similar. The Methanomicrobiales, in contrast to their supposed close relatives, are strictly anaerobic and active at redox potentials between –350 and –450 mV. They are not heterotrophic and in the chemolithotrophic process of making cell material from $H_2$ and $CO_2$, they produce methane ($CH_4$) using a number of unique energy-generating processes [6]. If the haloarchaea and their aerobic, heterotrophic photosynthetic characteristics were derived from a methanogenic ancestor [1], then either there is a need to find an explanation for this vast number of physiological changes or the SSU rRNA-derived phylogenetic position is incorrect and the haloarchaea are more closely related to some other group of prokaryote. An explanation of haloarchaeal evolution could therefore provide an insight into how this dramatically different biological system arose.

One explanation could be that simple point mutations have fuelled the extraordinary changes in the Haloarchaea, but the possibility arises that they came about by some alternative processes. Some composite phylogenies have suggested that the placing of the haloarchaea in the middle of the methanogens may be incorrect [7, 8]{Clarke, 2002 #39}{House, 2002 #40}. These results are based on the analysis of available phylogenies [7] or on the basis of shared orthologs [8]{Clarke, 2002 #39}.  In this scenario, the haloarchaea are placed as one of the earliest branching Archaea. The phenotypic characteristics of the haloarchaea are therefore easily interpreted as the retention of ancestral functions and their loss in other Archaea.  However, the placement of the haloarchaea as an early branching Archaeon may simply be the result of including genes with conflicting evolutionary histories {Brochier, 2004 #38}.

One such source of conflicting evolutionary history comes from Horizontal Gene Transfer (HGT), the mechanism and influence of which on the prokaryote tree of life has been extensively debated [9-14]. Generally, analyses of HGT have involved the identification of alien sequences either without reference to other genomes (e.g. the use of synonymous codon usage statistics) [15] or by searching for homologues in available databases or whole genomes (using string matching techniques like BLAST [16]). Both of these methods identify alien genes in the species in question, but in a phylogenetic context they only identify events that have occurred at the tips of a tree [17]. A third method used for the identification of HGT involves the comparison of a gene tree to a species tree. The advantage of this method is that it is possible to identify those genes that have been transferred as those that are "out of place" when compared to the species tree [18]. The problems with this method include the difficulties inherent in phylogeny reconstruction [19] and in distinguishing "bad" phylogenies from HGT [9].

In order to be as rigorous as possible we used a combination of these approaches when identifying possible HGTs. Firstly we identified archaeal specific gene families (in single copy) from completely annotated genomes to infer a robust (species) tree for the Archaea using total evidence and supertree approaches. We then analysed the haloarchaeal clade for ancestral inter-domain HGT transfers from Bacteria using a phylogenetic approach in order to identify genes that could potentially have led to the emergence of the haloarchaeal phenotype.

## Methods:

The amino acid sequences of 156 completely-sequenced genomes were retrieved from the COGENT database [20] and one (*Haloarcula marismortui*) from the NCBI. These genomes included 12 Eukarya, 17 Archaea and the remaining 128 represented a broad spectrum of available bacterial groups (see online supplementary information for a complete list). We performed all-against-all BLASTP searches with these genomes and identified gene families where all members were capable of identifying all other members of the family (and nothing else) using an e-value cut-off of $10^{-7}$. This approach is very conservative but it decreases the likelihood of using spliced genes and of inadvertently using paralogs for species tree inference.

## Construction of an archaeal species phylogeny:

**1) Taxonomic congruence analysis**

The subset of archaeal-specific single-gene families were used for the purposes of constructing an archaeal species phylogeny. We eliminated gene families where there was more than one member of the family in any given archaeal genome (multi-gene families) or where there was a member outside the Archaea. Two–hundred and twenty gene families were identified that met these criteria. Each of these families were aligned with ClustalW 1.81 [21] using the default settings. In order to test for the presence of a level of signal that is greater than randomness, 1,000 replicates of the Permutation Tail Probability (PTP) test were carried out on each alignment using PAUP* 4.0b10 [22], where the p-value cut-off was 0.001. Fifteen of the alignments failed this test so phylogenetic hypotheses were constructed using only the remaining 205 (these are available from the online supplementray information). The phylogenetic trees were constructed using Puzzleboot

(http://hades.biochem.dal.ca/Rogerlab/Software/software.html) and Tree-Puzzle [23]. A total of 100 bootstrap replicates were created for each of the 205 gene family alignments using Seqboot from the PHYLIP package [24]. These replicates were then used by Tree-Puzzle to create 100 distance matrices for each gene family using the JTT model [25]. The model of sequence evolution assumed there were two classes of sites – one being invariable and the other class being free to change. The rate variation across the variable sites was assumed to follow a gamma shape distribution that was approximated using a discrete approximation with eight categories of sites. Phylogenetic hypotheses were then constructed based upon these distance matrices using the neighbor-joining algorithm [26] implemented in the NEIGHBOR program of the PHYLIP package [24]. Agreement across these 100 trees was then summarised using a majority-rule consensus where only hypotheses of relationships with greater than 70% Bootstrap Proportion (BP) support were retained (the resulting 205 trees are available from the online supplementary information).

A phylogenetic supertree approach was then used in order to evaluate the level of agreement across these gene trees. All 205 gene trees were used as input for the Most Similar Supertree Algorithm (MSSA) [9] and the Matrix Representation using Parsimony (MRP) method as implemented in Clann [27]. One hundred bootstrap replicates were carried out using both these methods (see [28] for properties of the method).

**2) Total evidence analysis**

Thirty-seven of the 205 single gene families were universally distributed across all 17 archaeal taxa, and their aligned sequences were concatenated together for a Total Evidence (TE) analysis. The initial stage of the TE analysis consisted of classifying all the amino acid positions of the concatenated alignment into different rate categories. Tree-Puzzle [23] was used for this purpose assuming the JTT model [25] and with site rate heterogeneity approximated using a gamma shape distribution. Eight categories of sites were assumed to exist in this alignment and each of the sites of the alignment were placed into one of these eight rate categories. Using this classification, three alignments were constructed: the first contained all categories of sites; the second with category eight (the fastest-evolving) removed; and the third with categories seven and eight removed. Phylogenetic trees were constructed for each of these three alignments using two approaches. The first was a Bayesian posterior probability with Markov Chain Monte-Carlo (MCMC) approach (using the JTT model of amino acid substitution) as implemented in MrBayes [29]. The second approach used Puzzle-Boot and Tree-Puzzle as described earlier. The branch lengths on the tree were calculated from the complete alignment of the 17 universally distributed genes using PROTDIST and NEIGHBOR from the PHYLIP package. The three alignments are available from the online supplementary information.

## Identification of inter-Domain HGT events

### 1) Phylogenetic analysis of Haloarchaeal genes

All *Halobacterium NRC-1* and *H. marismortui* genes that had homologues in Archaea and Bacteria (as defined by the gene families constructed) were analysed for HGT events using phylogenetic reconstruction. Each gene family was aligned using the default settings in ClustalW 1.81 [21]. One hundred bootstrap replicates were then created for each alignment using SEQBOOT from the PHYLIP package [24] and distance matrices were estimated for each bootstrapped alignment using the JTT substitution model as implemented in PROTDIST in PHYLIP [24]. Phylogenetic trees were then reconstructed for each distance matrix using the Neighbor-joining algorithm as implemented in NEIGHBOR from the PHYLIP package [24]. Finally, only hypotheses of relationships with greater than 70% bootstrap proportion (BP) support (as calculated using CONSENSE from the PHYLIP package [24]) from all 100 bootstrap replicates were retained in the final tree. The trees were then categorised by eye according to the evidence of HGT from the phylogenetic analysis. The four categories as illustrated in Figure 1 are: Definite HGT (where a haloarchaeal gene arose from within a strongly supported bacterial clade), Likely HGT (where a haloarchaeal gene was placed with high BP support as a sister taxon to a bacterial clade), Unlikely HGT (Where there was no BP support for a definite phylogenetic position of a haloarchaeal gene), No HGT (Where the phylogenetic position of the haloarchaeal query gene was placed with strong support either within an archaeal group or as sister taxon to an archaeal group).

**2) Identification of HGT events in the haloarchaeal stem lineage.**

In the event that a sequence from *Halobacterium* NRC-1 and a sequence from *H. morismortui* were recovered as strongly-supported sister taxa, but both sequences were nested within a bacterial clade or placed as sister group to such a clade, this was taken as evidence for the acquisition of a gene in the halobacterium stem lineage.

## Results:

**Phylogenetic position of the haloarchaea within the Archaea.**

The concatenated alignment of the thirty-seven universally distributed genes used in the TE analysis had a total alignment length of 10,324 positions. Based on their rate heterogeneity, 1,648 amino acid positions were classified into category eight (the most rapidly-evolving) and 1,405 amino acid positions were classified into category seven.

A summary of the results of the six different total evidence (TE) phylogenetic analyses carried out are shown in Figure 2. Every branch of this tree received clade probability support of 1 from the MCMC analysis, strongly supporting the sister-group placement of the Halobacteriales and the Methanosarcinales. This indicates that the haloarchaea have a methanogenic ancestry. *Nanoarchaeum equitans* was used as an outgroup as phylogenetic trees derived from ribosomal proteins indicate that it is one of the earliest diverging groups of Archaea (it may be a deeply branching Euryarchaeote or Crenarchaeote, however neither of these positions would effect the results presented here) [30].

In the Tree-Puzzle analyses the support increased as the most rapidly evolving sites were stripped until when categories 7 and 8 were both stripped, every branch received greater than 95% BP support. This phylogenetic tree agrees with results from Ribosomal RNA gene phylogenies [4] and with respect to the position of the haloarchaea it disagrees with the results of gene content analyses [7, 8].

The taxonomic congruence (TC) analyses carried out on the 205 archaeal-specific gene trees also supported the tree in Figure 2. The gene trees were generated using a total of 64,709 aligned positions (with an average alignment length of 315 amino acids). The number of input trees with 4 to 17 taxa were as follows: 43, 33, 19, 17, 11, 6, 6, 8, 3, 4, 7, 4, 7, 37, respectively. When the underlying input trees were examined, 39% were identical to the best supertree pruned to the same taxa-set (i.e. completely compatible). This compares well with the 46% found to be completely compatible with the γ-proteobacterial phylogenetic supertree described previously using the same approach [9].

**Origin of archaeal genes**

Using a phylogenetic approach, we specifically looked for transfers into the haloarchaea from Bacteria. Paying attention to each haloarcheal genome separately and only to those genes for which we could make robust inferences of phylogenetic relationships, we could make a statement about 771 genes in *Halobacterium NRC-1*. A total of 277 of these genes (35.9%) were interpreted as definite inter-domain HGT, 56 (7.2%) as likely HGT and 438 (56.8%) were identified as definitely not HGT. In *H. marismortui* we could make statements about 1,194 genes. A total of 465 genes (38.9%) were interpreted as definite inter-domain HGT, 96 (8%) as likely inter-domain HGT and 633 (53%) as definitely not HGT. When these results are combined, 281 genes are present in both *Halobacterium NRC-1* and *H. marismortui*, are monophyletic with greater then 70% BP support and appear to be acquired from Bacteria. In total, our results indicate that in both haloarchaea, almost half the genes are not archaeal in origin, but bacterial. Although we were restricted by the limits of genome sampling and the uncertainty inherent in phylogenetic methods and our own desire to make conservative estimates, the number of genes involved can be considered large enough to be reasonable estimates of the phylogenetic affiliations of the entire dataset. We expect that in the fullness of time and with greater sampling, definite statements will be made about a larger proportion of the genes in these genomes. However, the signal, irrespective of the methods of analysis indicates that the haloarchaeal genomes are almost as much bacterial as archaeal.

**Functions of ancestrally acquired genes in haloarchaea**

Of those definite and likely HGT events identified, 281 occurred prior to the separation of *Halobacterium NRC-1* and *H. marismortui*. The largest functional category in this group were the transporters making up 21% of all the genes for which we could assign a function, followed by energy metabolism making up 11.5% (Table 1). The next most highly represented functional categories were regulation (7.5%), cell-envelope components (5.3%) and cellular processes (4.4%). The nutrient uptake genes acquired through ancient HGT events include ABC transporters for sugars and ribose, glycerol 3-phosphate, spermidine/putrescine, phosphate, zinc and copper. Additionally, Iron permease and $Na^+/H^+$ antiporters have also been acquired through ancient HGT events. Figure 3 shows one example of some interacting metabolic pathways that contain ancestral HGT genes, these include 6 out of the 8 genes possessed by *H. marismortui* for the anaerobic reoxidation of pyruvate, 3 out of the 16 genes for glycolysis/gluconeogenesis and 3 out of the 10 genes involved in the citrate cycle. Considering the conservative approach we have taken at every step of our analysis we expect that the numbers of horizontally transferred genes into each of these metabolic pathways is actually higher than reported here. Nonetheless even just taking our results at face value demonstrates the contribution from the bacterial domain in these organisms.

**Discussion:**

It has been suggested previously that the genome sequence of *Halobacterium NRC-1* has a significant bacterial character [1, 31], but the full extent of the situation has not been identified. Furthermore, in the absence of a species phylogeny for the Archaea which has been constructed from those genes known to exist only in the Archaea, its interpretation has been problematic [8].

We have shown that using Archaea-specific genes, the phylogenetic placement of the haloarchaea as derived methanogens is correct. This contradicts the results of a previous analysis of shared orthologs [8]. As expected, their position as topological neighbors of the Methanosarcinales and as a relatively late-diverging lineage is strongly supported in all analyses, indicating that our analyses are not effected by systematic biases that may cause the haloarchaea to be "deep-branching". From the HGT analysis we estimate that up to 47% of *H. marismortui* and 43% of *Halobacterium NRC-1* genes are bacterial in origin. This is a clear and consistent signal. The phylogenetic placement of the haloarchaea as early diverging archaea based on shared orthologs [8] seems most likely an artefact of this significant bacterial contribution.

Because of the strong support for the placement of the haloarchaea as a late diverging euryarchaeote from the taxonomic congruence and total evidence analyses, it seems very unparsimonious to suggest that the haloarchaeal genes that only have bacterial homologs are ancestrally derived. We would have to invoke additional large-scale independent losses of genes in the other Archaea. The same reasoning holds true for those genes with greatest similarity to bacterial genes and with lower levels of similarity to archaeal genes. This leaves HGT as the most likely cause for the presence of these genes.

The scale of the change necessary for a $CO_2$-reducing chemolithotroph to develop into an aerobic autophosphorylating heterotroph is enormous, and the genes from the ancestral HGT events are likely to be the haloarchaea-making genes. For example, heterotrophy would require the ability to take up organic molecules (like amino acids, vitamins, lipids or glucose) and bacterial ABC transporters (which make up nearly half of all the transporters involved in the ancestral HGT events) are designed for these purposes. Another obstacle to heterotrophy would be energy production and we find that the second largest group of genes from ancestral HGT events are involved in energy metabolism (Table 1). Other requirements for heterotrophy are met with bacterial cell envelope components, and cellular processes acquired through ancient HGT. While we do find a large number of haloarchaeal genes from key metabolic pathways originated from the Bacteria (Table 1), they do not makeup the majority of the genes in these pathways. This may be because of our desire to be as conservative as possible at each step of the analysis or simply because we cannot identify homologs in any other genome for over 40% of the genes in both of the haloarchaea. A third possibility however is that these key metabolic processes only require minor changes to make use of the molecules provided by the new transport systems incorporated into the organism. Future analyses may clarify this point.

The scale of the transfers from the bacteria to the haloarchaea suggests an evolutionary scenario similar to that proposed by the "hydrogen hypothesis" {Martin, 1998 #43} whereby a methanogen evolves into a heterotroph by endosymbiosis with a eubacteria. In that instance, the gene transfers occurred by endosymbiotic gene transfer, resulting in a genome comprised of a mosaic of archaeal and eubacterial genes {Esser, 2004 #44}. As attractive as it would be to suggest our results support this scenario, we have not been able to pinpoint a single bacterial donor of the genes horizontally transferred into the archaea, even though there was a consistent phylogenetic signal from the bacterial domain. This may be because there wasn't a single donor, but also it could be due to the sparseness of current sampling of bacterial genomes or the inadequacy of sequence-based methods of reconstructing relationships within the Bacteria. More likely though is that the acquisition of the genes occurred in a cascade of HGT events. Recent studies have suggested that there is elevated rates of both HGT and recombination in the haloarchaea [32, 33], however the level of inter-domain HGT reported here is unprecedented. Interestingly though, the tree in figure 2 suggests that the transition from methanogen may have occurred at least 3 independent times, leading to Thermoplasmatales, Archaeoglobales and Halobacteriales, and merits further study. Indeed, sequencing the genome of *Methanosarcina mazei* (which belongs to the sister group to the haloarchaea (Figure 2)) revealed putative horizontal transfers from the bacteria that may have had widespread effects on its metabolic capability. Perhaps such drastic physiological changes are easily accepted in these groups; certainly the major organisational and content differences between the two closely related haloarchaeal genomes {Baliga, 2004 #41} suggest this may be the case.

We may also speculate that when we find homologs of the 43% of the *H. marismortui* and 44% of the *Halobacterium NRC-1* genes that are without relatives in the currently annotated genomes, the majority will probably be operational genes.

This is because we are unlikely to discover entirely new information-processing schemes. Given that informational genes in these genomes are overwhelmingly of archaeal origin and operational genes have a much stronger tendency to be bacterial, the final count for the haloarchaea could be that their genomes have a greater number of bacterial genes than archaeal genes.

**Acknowledgements**

The Higher Education Authority Programme for research in Third Level Institutes Cycle III funded this work. C.J.C. and J.O.M. both designed the project and wrote the manuscript. C.J.C. carried out the data analysis. The online supplementary information is available at:

http://bioinf.nuim.ie/supplementary/archaea/ . Correspondence and requests for materials should be addressed to James McInerney, email:

james.o.mcinerney@nuim.ie.

## Legends:

### Figure 1

Determining the origin of haloarchaeal genes. Following the retention of only the most highly supported (greater than 70% BP support) hypotheses of relationships, each of the trees containing a haloarchaeal gene was examined by eye to determine it's origin. The grey branches represent the four broad classifications of the gene position used to determine its likelihood of having undergone a HGT. A) Definite HGT: The haloarchaeal gene was placed within a highly supported group of Bacteria separate to other Archaea. B) Likely HGT: The haloarchaeal gene was placed with high support as a sister group to a Bacterial clade. C & D) Not HGT: The haloarchaeal gene was placed as a sister group to an Archaeal or was placed within an Archaeal clade. The only classification not illustrated is Unlikely HGT: when the haloarchaeal gene had low bootstrap support for its placement in the tree.

### Figure 2

The hypotheses of archaeal relationships of as reconstructed using all six total evidence (TE) approaches. All three analyses using MrBayes found clade probability values of 1 for every branch. With the Puzzle analysis, as the fastest sites were stripped the support for the tree increased until there was greater then 95% BP support for all internal branches. The tree also agreed with those strongly supported clades from the two Taxonomic Congruence (TC) approaches.

**Figure 3**

An example of interacting metabolic pathways for which ancestral haloarchaeal HGT genes have been found. The central pathway is the anaerobic reoxidation of pyruvate (ARP). The conversion of phosphoenolpyruvate to pyruvate is the last step of glycolysis. The pathways in blue are those that interact with the products of glycolysis or ARP that contain genes that have been identified to have been acquired through ancient HGT. The fractions in the blue boxes are the number of definite ancient HGT genes present, out of the total number of genes that *H. marismortui* possesses from the pathway. The circles represent products and the rectangles connecting them indicate the enzymes that produce those products. The enzymes in yellow are the genes from ARP that have been identified as having arisen through ancient HGT. The enzymes in yellow are as follows: *2.7.1.40:* pyruvate kinase; *1.2.4.1:* pyruvate dehydrogenase E1 component alpha subunit; *1.8.1.4:* dihydrolipoamide dehydrogenase; *1.1.1.1* and *1.1.1.2:* alcohol dehydrogenase; *1.2.1.3:* aldehyde dehydrogenase.

**Table 1**

The functional categories of the genes acquired prior to the separation of *Halobacterium NRC-1* and *H. marismortui* (the ancestral haloarchaeal HGT events).

Figure 1:

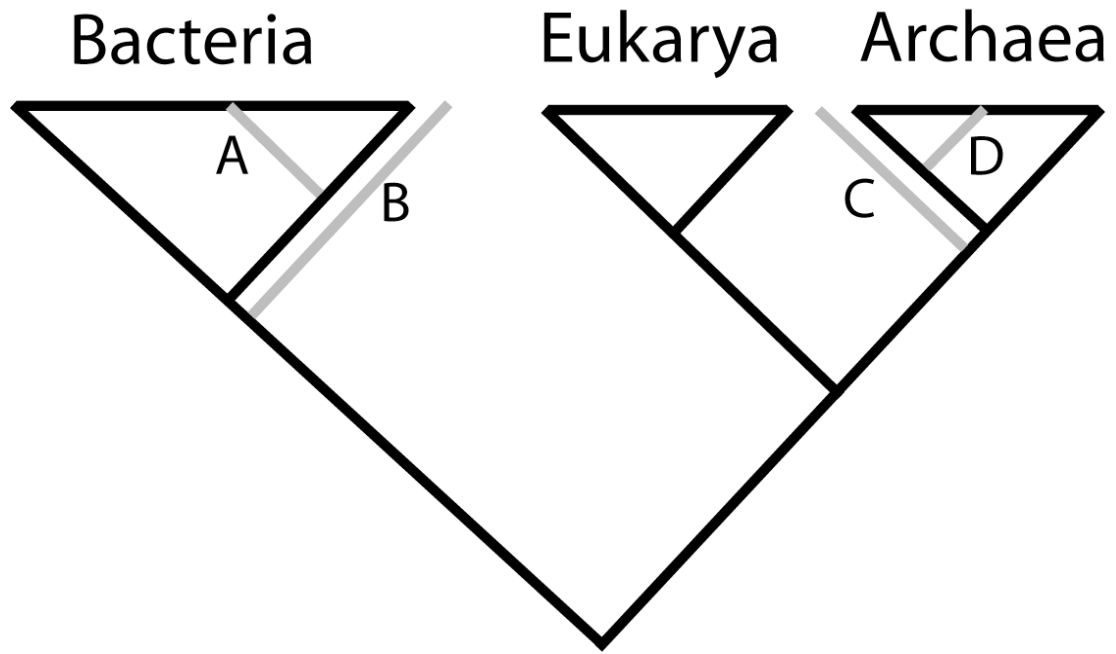

Figure 2:

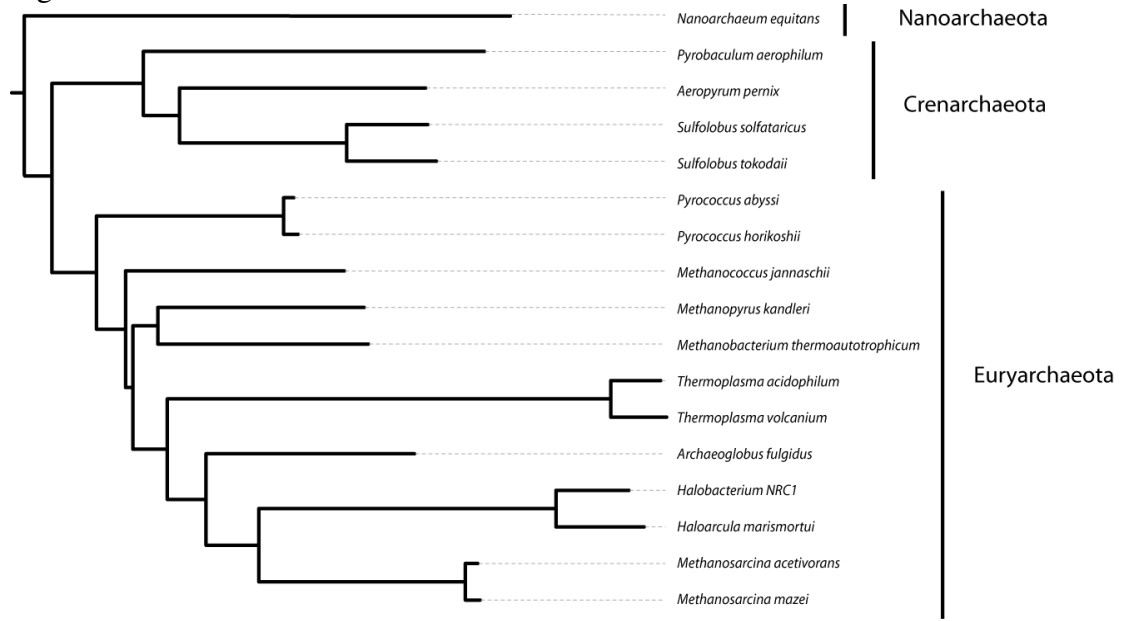

Figure 3:

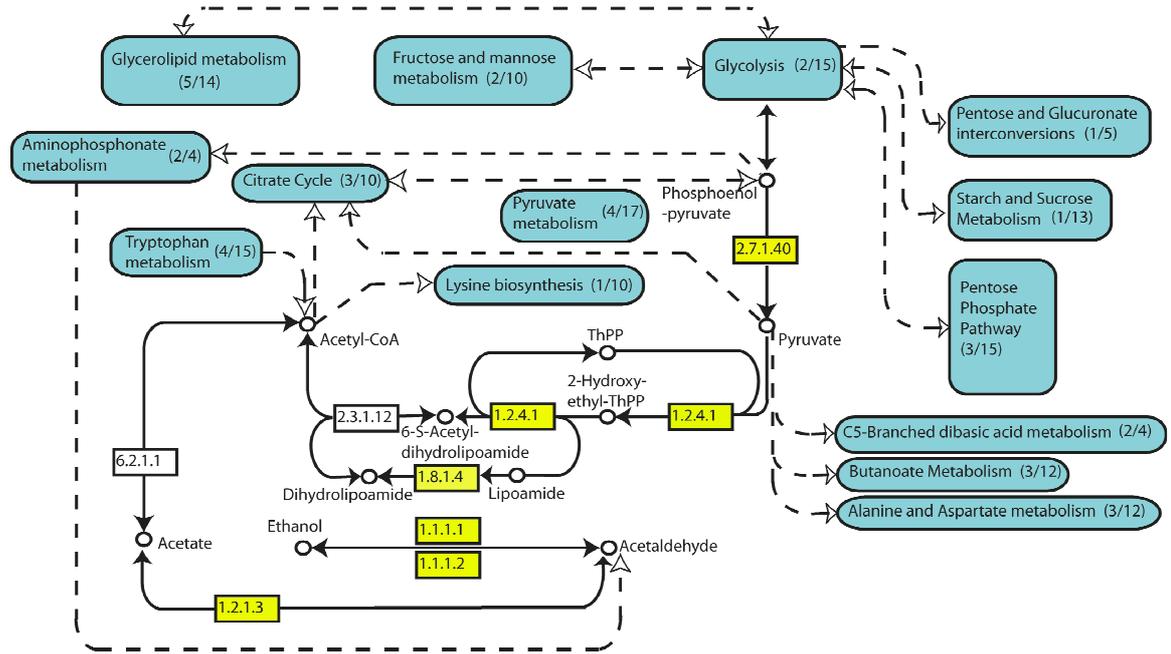

*Table 1:*

| Functional Classification | Number of genes |
|---|---|
| Amino Acid metabolism | 9 |
| Nucleotide metabolism | 6 |
| Cofactor metabolism | 4 |
| Energy Metabolism | 26 |
| Cell envelope components | 12 |
| Transport | 49 |
| Cellular processes | 10 |
| DNA replication, repair, recombintion | 6 |
| Transcription | 2 |
| Regulation | 17 |
| Translation | 5 |
| Miscellaneous | 80 |
| Hypothetical | 55 |

**References**

1. Ng WV, Kennedy SP, Mahairas GG, Berquist B, Pan M, Shukla HD, Lasky SR, Baliga NS, Thorsson V, Sbrogna J et al: **Genome sequence of Halobacterium species NRC-1**. *Proc Natl Acad Sci U S A* 2000, **97**(22):12176-12181.
2. Magrum LJ, Luehrsen KR, Woese CR: **Are extreme halophiles actually "bacteria"?** *J Mol Evol* 1978, **11**(1):1-8.
3. Soppa J: **From genomes to function: haloarchaea as model organisms**. *Microbiology* 2006, **152**(Pt 3):585-590.
4. Matte-Tailliez O, Brochier C, Forterre P, Philippe H: **Archaeal phylogeny based on ribosomal proteins**. *Molecular Biology and Evolution* 2002, **19**(5):631-639.
5. Ciccarelli FD, Doerks T, von Mering C, Creevey CJ, Snel B, Bork P: **Toward automatic reconstruction of a highly resolved tree of life**. *Science* 2006, **311**(5765):1283-1287.
6. Vogels GD, van der Drift C, Stumm CK, Keltjens JT, Zwart KB: **Methanogenesis: surprising molecules, microorganisms and ecosystems**. *Antonie Van Leeuwenhoek* 1984, **50**(5-6):557-567.
7. Wolf YI, Rogozin IB, Grishin NV, Koonin EV: **Genome trees and the Tree of Life**. *Trends in Genetics* 2002, **18**(9):472-479.
8. Korbel JO, Snel B, Huynen MA, Bork P: **SHOT: a web server for the construction of genome phylogenies**. *Trends in Genetics* 2002, **18**(3):158-162.
9. Creevey CJ, Fitzpatrick DA, Philip GK, Kinsella RJ, O'Connell MJ, Pentony MM, Travers SA, Wilkinson M, McInerney JO: **Does a tree-like phylogeny only exist at the tips in the prokaryotes?** *Proc R Soc Lond B Biol Sci* 2004, **271**(1557):2551-2558.
10. Kurland CG, Canback B, Berg OG: **Horizontal gene transfer: A critical view**. *Proc Natl Acad Sci U S A* 2003, **100**(17):9658-9662.
11. Doolittle WF: **Phylogenetic classification and the universal tree**. *Science* 1999, **284**:2124-2129.
12. Woese CR: **Interpreting the universal phylogenetic tree**. *Proceedings of the National Academy of Sciences of the United States of America* 2000, **97**(15):8392-8396.
13. Sorek R, Zhu Y, Creevey CJ, Francino MP, Bork P, Rubin EM: **Genome-wide experimental determination of barriers to horizontal gene transfer**. *Science* 2007, **318**(5855):1449-1452.
14. Dagan T, Martin W: **Ancestral genome sizes specify the minimum rate of lateral gene transfer during prokaryote evolution**. *Proc Natl Acad Sci U S A* 2007, **104**(3):870-875.
15. Lawrence JG, Ochman H: **Molecular archaeology of the Escherichia coli genome**. *Proc Natl Acad Sci U S A* 1998, **95**(16):9413-9417.
16. Altschul SF, Madden TL, Schaffer AA, Zhang J, Zhang Z, Miller W, Lipman DJ: **Gapped BLAST and PSI-BLAST: a new generation of protein database search programs**. *Nucleic Acids Research* 1997, **25**:3389-3402.
17. Poptsova MS, Gogarten JP: **The power of phylogenetic approaches to detect horizontally transferred genes**. *BMC Evol Biol* 2007, **7**:45.

18. Kinsella RJ, McInerney JO: **Eukaryotic genes in Mycobacterium tuberculosis? Possible alternative explanations**. *Trends Genet* 2003, **19**(12):687-689.
19. Felsenstein J: **Inferring Phylogenies**. Sunderland, Massachusetts: Sinauer Associates, Inc; 2003.
20. Janssen P, Enright AJ, Audit B, Cases I, Goldovsky L, Harte N, Kunin V, Ouzounis CA: **COmplete GENome Tracking (COGENT): a flexible data environment for computational genomics**. *Bioinformatics* 2003, **19**(11):1451-1452.
21. Chenna R, Sugawara H, Koike T, Lopez R, Gibson TJ, Higgins DG, Thompson JD: **Multiple sequence alignment with the clustal series of programs**. *Nucleic Acids Res* 2003, **31**(13):3497-3500.
22. Swofford DL: **PAUP*. Phylogenetic Analysis Using Parsimony (*and other methods). Version 4**: Sinauer Associates, Sunderland, Massachusetts.; 2002.
23. Schmidt HA, Strimmer K, Vingron M, von Haeseler A: **TREE-PUZZLE: maximum likelihood phylogenetic analysis using quartets and parallel computing**. *Bioinformatics* 2002, **18**(3):502-504.
24. Felsenstein J: **Phylip: Phylogeny Inference package**. In., 3.6 edn. Seattle: Distributed by author; 1993.
25. Jones DT, Taylor WR, Thornton JM: **The rapid generation of mutation data matrices from protein sequences**. *Comput Appl Biosci* 1992, **8**(3):275-282.
26. Saitou N, Nei M: **The neighbor-joining method: a new method for reconstructing phylogenetic trees**. *Mol Biol Evol* 1987, **4**(4):406-425.
27. Creevey CJ, McInerney JO: **Clann: investigating phylogenetic information through supertree analyses**. *Bioinformatics* 2005, **21**(3):390-392.
28. Wilkinson M, Cotton JA, Creevey C, Eulenstein O, Harris SR, Lapointe FJ, Levasseur C, McInerney JO, Pisani D, Thorley JL: **The shape of supertrees to come: tree shape related properties of fourteen supertree methods**. *Syst Biol* 2005, **54**(3):419-431.
29. Huelsenbeck JP, Ronquist F: **MRBAYES: Bayesian inference of phylogenetic trees**. *Bioinformatics* 2001, **17**(8):754-755.
30. Waters E, Hohn MJ, Ahel I, Graham DE, Adams MD, Barnstead M, Beeson KY, Bibbs L, Bolanos R, Keller M *et al*: **The genome of Nanoarchaeum equitans: insights into early archaeal evolution and derived parasitism**. *Proc Natl Acad Sci U S A* 2003, **100**(22):12984-12988.
31. Kennedy SP, Ng WV, Salzberg SL, Hood L, DasSarma S: **Understanding the adaptation of Halobacterium species NRC-1 to its extreme environment through computational analysis of its genome sequence**. *Genome Res* 2001, **11**(10):1641-1650.
32. Papke RT, Koenig JE, Rodriguez-Valera F, Doolittle WF: **Frequent recombination in a saltern population of Halorubrum**. *Science* 2004, **306**(5703):1928-1929.
33. Zhaxybayeva O, Gogarten JP: **Bootstrap, Bayesian probability and maximum likelihood mapping: exploring new tools for comparative genome analyses**. *BMC Genomics* 2002, **3**(1):4.